\documentclass[times,astrobib,amssymb,usenatbib]{mn2e}
\usepackage{graphicx}
\usepackage{txfonts}
\usepackage{natbib}
\usepackage{psfig}

\begin{document}

\newcommand{\lapp}{\mbox{\raisebox{-0.3em}{$\stackrel{\textstyle <}{\sim}$}}}
\newcommand{\gapp}{\mbox{\raisebox{-0.3em}{$\stackrel{\textstyle >}{\sim}$}}}
\newcommand{\be}{\begin{equation}}
\newcommand{\en}{\end{equation}}
\newcommand{\di}{\displaystyle}
\def\tworule{\noalign{\medskip\hrule\smallskip\hrule\medskip}} 
\def\onerule{\noalign{\medskip\hrule\medskip}} 
\def\bl{\par\vskip 12pt\noindent}
\def\bll{\par\vskip 24pt\noindent}
\def\blll{\par\vskip 36pt\noindent}
\def\rot{\mathop{\rm rot}\nolimits}
\def\alf{$\alpha$}
\def\refff{\leftskip20pt\parindent-20pt\parskip4pt}
\def\zabs{$z_{\rm abs}$}
\def\zem{$z_{\rm em}$~}
\def\lya{Ly$\alpha$ }
\def\lyb{Ly$\beta$ }
\def\hi{H\,{\sc i}~}
\def\mgii{Mg\,{\sc ii}~}
\def\feiia{Fe\,{\sc ii}$\lambda$2600}
\def\mgia{Mg\,{\sc i}$\lambda$2852}
\def\mgiia{Mg\,{\sc ii}$\lambda$2796}
\def\mgiib{Mg\,{\sc ii}$\lambda$2803}
\def\mgiiab{Mg\,{\sc ii}$\lambda\lambda$2796,2803}
\def\wobs{$w_{\rm obs}$}
\def\kms{km~s$^{-1}$}

\title{SDSS J092712.64+294344.0: recoiling black hole or merging galaxies$?$}
\author[M. Vivek, R. Srianand, P. Noterdaeme, V. Mohan and V. C. Kuriakose]{M. Vivek$^{1,2}$\thanks{E-mail: vivekm@iucaa.ernet.in}
R. Srianand$^{2}$, P. Noterdaeme$^2$, V. Mohan$^2$ \& V. C. Kuriakose$^1$
\\
$^{1}$ Cochin University of Science and Technology, Kochi 682\,022, India\\ 
$^{2}$ Inter-University Centre for Astronomy and Astrophysics, Post Bag 4,  Ganeshkhind, Pune 411\,007, India \\}

\date{Accepted. Received; in original form }
\pagerange{\pageref{firstpage}--\pageref{lastpage}} \pubyear{2009}

\maketitle

\label{firstpage}

\begin{abstract}
We report long-slit spectroscopic observations of SDSS\,J092712$+$294344 carried-out at the recently 
commissioned 2\,m telescope in IUCAA Girawali Observatory, India. This AGN-like source is known to 
feature three sets of emission lines at \zem = 0.6972, 0.7020 and 0.7128. Different scenarios such as
a recoiling black hole after asymmetric emission of gravitational waves, binary black holes and possible 
merging systems are proposed for this object. 
We test these scenarios by comparing our spectra with that from the Sloan Digital Sky Survey 
(SDSS), obtained 4 years prior to our observations.
Comparing the redshifts of [O\,{\sc iii}]$\lambda\lambda$4960,5008  
we put a 3$\sigma$ limit on the relative acceleration to be less than 
32~\kms\,yr$^{-1}$ between different emitting regions. 
Using the 2D spectra obtained at different position angles we show that the 
[O\,{\sc iii}]$\lambda5008$ line from the \zem = 0.7128 component is extended 
beyond the spectral point spread
function. We infer the linear extent of this line emitting region is $\sim$ 8 kpc. 
We also find a tentative evidence for an offset between the centroid of the [O\,{\sc iii}]$\lambda$5008 
line at \zem = 0.7128 and the QSO trace when the slit is aligned at a position angle of 
299$^\circ$. This corresponds to the \zem = 0.7128 system being at an 
impact parameter of $\sim$1~kpc with respect to the \zem = 0.6972 in the north west direction.
Based on our observations we conclude that the binary black hole model is most unlikely.
The spatial extent and the sizes are consistent with both black hole recoil and
merging scenarios.
\end{abstract}

\section{Introduction}

The availability of several thousands of QSO spectra in the Sloan Digital Sky Survey (SDSS) database 
has allowed astronomers to find various interesting and peculiar AGNs. In particular, the discovery of unresolved
point sources with two sets of emission lines that are powered by AGN-like continuum sources
[SDSS\,J092712$+$294344 at $z = 0.713$ \citep{Komossa08}, SDSS\,J153636$+$044127 at $z = 0.38$ 
\citep{Boroson09} and SDSS\,J105041$+$345631 at $z = 0.272$ \citep{Shields09b}] has opened 
up possibilities to study recoiling black holes and/or binary inspiralling super-massive black holes. 
In this paper we concentrate on the first object (i.e. SDSS\,J092712.64$+$294344.0, hereafter J0927$+$2943).
J0927$+$2943 is an unusual quasar with $z=0.713$, identified by \citet{Komossa08} 
during their search for active galactic nuclei with high [O\,{\sc iii}] velocity shifts. 
There are two systems of emission lines identified in the SDSS spectrum  with 
a velocity separation of about 2650~\kms. One is referred as 
`red' (with $z_r$ = 0.71279) and other as `blue' (with $z_b$ = 0.69713).  The red system 
consists of narrow emission lines (NELs) of [O\,{\sc iii}]$\lambda$5008, [O\,{\sc ii}]$\lambda$3727, 
[Ne\,{\sc iii}]$\lambda$3869,  [Ne\,{\sc v}]$\lambda$3426 and narrow Balmer lines. The blue system shows 
classical Balmer and Mg\,{\sc ii}  broad emission lines (BELs), plus unusually broad NELs.   
The line ratios indicate AGN-like excitation in both systems. \citet{Shields09a} reobserved 
this object with the Hobby-Eberly Telescope (HET) and reported a third redshifted set of narrow 
lines at $z \sim 0.7020$. They also put a bound on the line of sight acceleration between the red 
and blue systems (i.e a 3$\sigma$ limit of $dv/dt\le 24$ \kms yr $^{-1}$).

Simulations of binary black hole mergers predict large recoil velocities (kicks) of the final 
merged black hole resulting from anisotropic emission of gravitational radiation 
\citep[see for example,][]{Campanelli07a, Campanelli07b, Dain08, Gonzalez07, Loeb07, Tichy07}. 
In the discovery paper, \citeauthor{Komossa08} proposed J0927$+$2943 as a possible candidate for a 
super-massive black hole (with $M\sim10^{8.8}M_{\sun}$) ejected at high speed from the host galactic 
nucleus by gravitational radiation recoil. However, \citet{Dotti09} and \citet{Bogdanovic09} have 
proposed an alternate hypothesis in which the observed configuration of emission 
lines originate from binary black holes. The main features of this model are the prediction of a 
detectable acceleration over a time-scale of years and sub-parsec scale sizes for the emitting regions. 
At the same time, \citet{Heckman09} proposed that J0927$+$2943 could be a high
redshift analog of NGC\,1275 (also known as 3C\,84 and Perseus A) where two sets of redshifted
emission lines are seen due to the interactions between two galaxies in a cluster centre \citep{Conselice01}. 
{On similar lines, \citet{Shields09a} have proposed the superposition hypothesis based on the
third redshifted emission line component and a possible presence of substantial cluster apparently
containing J0927$+$2943. Subsequent detailed multiband photometric studies do not
substantiate the presence of a massive cluster at the redshift of  J0927$+$2943  \citep[see][]{Decarli09}.
However, it is still possible that J0927$+$2943 is part of a massive cluster with low luminous
matter to dark matter mass ratio.
}

The absence of
observable change in the redshifts of the emission lines and/or a clear proof of the line emitting gas spread over
kilo-parsec scales will clearly challenge the binary black hole scenario and favour the other two alternatives.
This paper aims at constraining both the acceleration and the spatial extent of the emitting regions using 
long-slit spectroscopy observations. 

\begin{table}
 \centering
\caption{Log of IGO observations}
 \begin{tabular}{l c c}
\hline
\multicolumn{1}{c}{Date}&Total exposure time$^1$ &Position Angle$^2$\\
    &\multicolumn{1}{c}{ (min)}                  &\multicolumn{1}{c}{(degrees)}\\
\hline
01/12/2008 & 120 &  66 \\
01/12/2008 & 90  & 202 \\
01/12/2008 & 90  & 299 \\
31/01/2009 & 120 & 100 \\
\hline
\end{tabular}
\begin{flushleft}
$^1$ Individual exposures are of 30 min duration.\\
$^2$ Angle measured in the clockwise direction with respect to south. In IGO
images, south is at the top and west is at the right hand side. 
\end{flushleft}
\label{logtable}
\end{table}

\section{Observations and Data Reduction}

We observed J0927$+$2943 using the 2 metre telescope at IUCAA Girawali Observatory (IGO) in India near Pune. 
Long-slit spectra covering the wavelength range 3700 to 9200~{\AA} were obtained using the GR5 grism of 
the IUCAA Faint Object Spectrograph (IFOSC) and a slit width of 1.5''. A typical seeing of 1.2 to 1.3'' was
measured from FWHM of the images taken during the nights.
The observations were carried out on 
01/12/2008 and 31/01/2009 for 4 different position angles of the slit. The detailed log of our 
observations is summarised in Table~\ref{logtable}. The raw CCD frames were cleaned using standard
IRAF{\footnote {IRAF is distributed by the National Optical Astronomy Observatories,
which are operated by the Association of Universities for Research
in Astronomy, Inc., under cooperative agreement with the National
Science Foundation.}} procedures. We use halogen flats for flat fielding the frames. 
Since at $\lambda>7000${\AA} simple flat fielding does not remove the fringes, the QSO was 
moved along the slit for different exposures of a same position
angle. We subsequently removed fringing by subtracting one frame from the other taken on the same night. 
The same procedure was applied to standard stars as well.

\begin{figure}
\centering
\psfig{figure=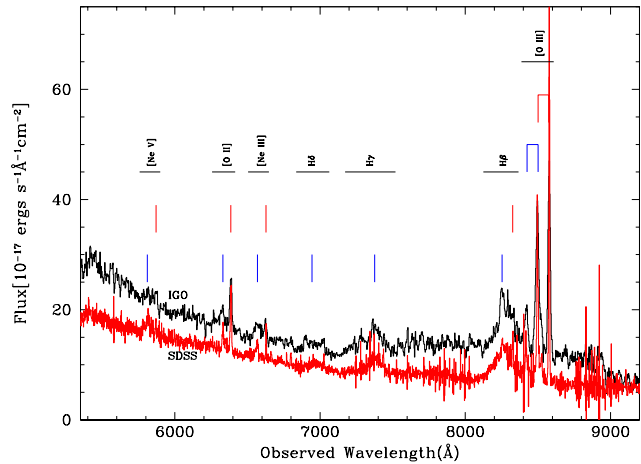,width=1.05\linewidth,angle=270}
\caption{Comparison of the spectra of SDSS J0927$+$2943 obtained with IGO and SDSS. Ticks
mark the locations of the emission lines from the red and blue systems. 
} 
\label{fig1}
\end{figure}

We then extracted the one dimensional spectrum from
individual frames using the IRAF task ``doslit''. 
Wavelength calibration of the spectra was performed using
Helium Neon lamps. 
Wavelengths were converted to the vacuum-heliocentric rest frame and individual 
spectra were scaled within a sliding window and coadded using 1/$\sigma^2$ weighting 
in each pixel. The error spectrum 
was computed taking into account proper error propagation during the combining process. The achieved 
spectral resolution is R~$\sim$~300 and the continuum signal-to-noise ratio in the combined spectrum 
varies between 20 and 40 per pixel. The final combined 1D spectrum together with the SDSS spectrum 
are shown in Fig.~\ref{fig1}.

In order to extract the spatial information from our observations, 
we model the observed individual 2D 
spectra in the wavelength range 8100-8700 {\AA} with the sum of the QSO continuum and 
the [O\,{\sc iii}] emission lines. The aim is to study the centroid shift of the 
emission lines with respect to the centre of the trace and the extent (FWHM) of the 
emission lines compared to that of the trace in the regions free from emission lines.
We subtract the sky from each science frame using the mean sky spectrum 
extracted from the spatial bins on either side of the QSO trace.

\section{Analysis and Results}

\begin{figure}
\centering
\psfig{figure=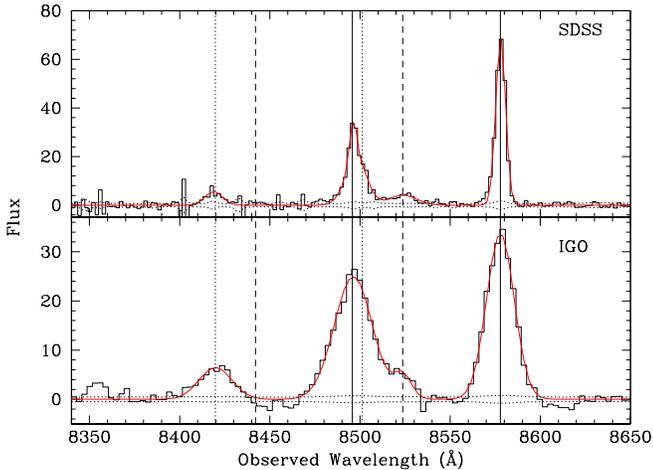,width=1.05\linewidth,angle=270}
\caption{Comparison of the spectra of SDSS J0927$+$2943 obtained with IGO and SDSS. 
Best fit Gaussians are over-plotted. The solid, dotted and dashed vertical lines
mark the locations of the different [O\,{\sc iii}] lines from the red, blue and 
the third system (at \zem = 0.7020).} 
\label{fig2}
\end{figure}

\begin{table}
\caption{Redshift measurements and velocity shifts}
\centering
\begin{tabular}{rccc}
\hline
\multicolumn{1}{c}{Line} & \multicolumn{2}{c}{Redshift} & $\Delta v$ \\
     & SDSS & IGO                   & (\kms) \\
\hline
[O\,{\sc iii}]$\lambda$5008 & 0.71275(1) & 0.71275(2) &   0$\pm$4 \\

[O\,{\sc ii}]$\lambda$3727  & 0.71285(5) & 0.71268(6) & 30$\pm$14 \\

[O\,{\sc iii}]$\lambda$5008 & 0.70203(14) & 0.70203(10) &  0$\pm$30 \\

[O\,{\sc iii}]$\lambda$4960 & 0.69724(10) & 0.69745(9) & 37$\pm$25 \\

\hline
\end{tabular}
\label{tab2}
\end{table}

The SDSS observations were carried out on January 19$^{\rm th}$, 2005. Our observations were taken after
a time interval of $\sim$4 years. In the rest frame of the object, this corresponds to an elapsed
time of 2.35 years. Even though our spectrum covers a wide wavelength range we mainly concentrate on
the [O\,{\sc iii}] emission lines as they are the strongest ones seen in the spectrum.
We fit the [O\,{\sc iii}] lines with Gaussians using standard $\chi^2$ minimisation techniques. The 
Gaussian fits to IGO and SDSS 1D-spectra are shown in Fig.~\ref{fig2} and the corresponding redshifts 
are given in Table~\ref{tab2}. For individual lines our 
measurement of the redshift matches well with that from the SDSS spectrum within errors.
Using [O\,{\sc iii}]$\lambda$5008 of the red system and [O\,{\sc iii}]$\lambda$4960 line
of the blue system we estimate the 3$\sigma$ limit on the acceleration at the redshift
of the QSO to be less than 32~\kms\,yr$^{-1}$. This is also consistent with the redshift of the red
component measured using the [O\,{\sc ii}]$\lambda$3727 line.

Note that this is very much comparable to the constraint
obtained by \citet{Shields09a} using [O\,{\sc ii}] and [Ne\,{\sc iii}] lines. However, it seems that
\citeauthor{Shields09a}'s values are not computed for the rest-frame elapsed time at the redshift of the QSO and 
their actual value may be higher by a factor 1.7 (i.e. a 3$\sigma$ limit of 41 \kms yr$^{-1}$).
Note that the constraint we get is a factor 3 less than the 
acceleration predicted by \citet{Bogdanovic09}. Our good S/N spectrum also confirms the third redshift found by \citet{Shields09a} at
\zem = 0.7028$\pm$0.0002.  Using the SDSS spectrum we find that emission from this system 
does not show any detectable acceleration. 

{\citet{Shields09a}, suggested that the presence of stellar Calcium H and K lines at redshift, $z_r$,
would undermine the recoil hypothesis. These lines are not detected in our IGO spectrum. We place
a 3$\sigma$ upper limit on the rest equivalent width of 0.4~{\AA} for the Ca~{\sc ii}$\lambda$3934 line.
However, this is not stringent enough to detect this line with the equivalent width ($\sim$ 0.2 \AA) 
as seen in the SDSS QSO composite spectrum \citep{VandenBerk01}.
}

\begin{table*}
\caption{Results from the 2D Gaussian fits to the [O\,{\sc iii}] lines}
\begin{tabular}{rccccc}
\hline
Position    & \multicolumn{3}{c}{FWHM (arcsec)} & \multicolumn{2}{c}{Shift (arcsec)$^3$} \\
Angle ($^{\circ}$) & Trace & L${_b}^1$ &  L${_r}^2$   & L$_b$    &  L$_r$\\
\hline
66  & 1.31(0.04) &   1.55(0.09) &  1.75(0.07) &  $+$0.06(0.13) & $+$0.11(0.04)\\
100 & 1.65(0.23) &   1.72(0.45) &  2.10(0.09) &  $+$0.00(0.08)  & $+$0.13(0.04)\\
202 & 1.29(0.04) &   1.21(0.09) &  1.70(0.10) &  $-$0.02(0.03) &  $+$0.05(0.04)\\
299 & 1.30(0.03) &   1.36(0.20) &  1.80(0.08) &  $-$0.10(0.05) &  $-$0.19(0.02)\\
\hline
\end{tabular}
\begin{flushleft}
$^1$ [O\,{\sc iii}]$\lambda$4960 blue system  \\
$^2$ [O\,{\sc iii}]$\lambda$5008 red system \\
$^3$ In the ccd image 1 pixel corresponds to 0.34''.\\
\end{flushleft}
\label{table1}
\end{table*}

\begin{figure}
\centering
\begin{tabular}{c}
\includegraphics[angle=-270, width=0.95\linewidth, bb=190 90 303 750,clip=true]{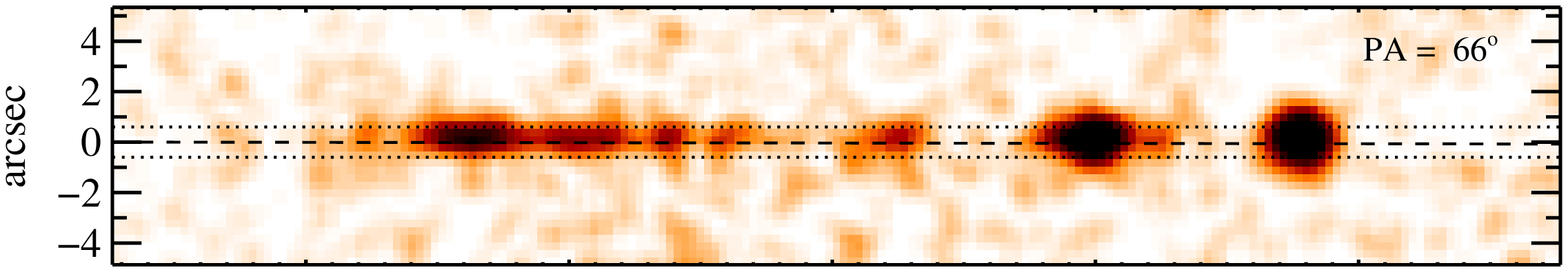}\\
\includegraphics[angle=-270, width=0.95\linewidth, bb=190 90 303 750,clip=true]{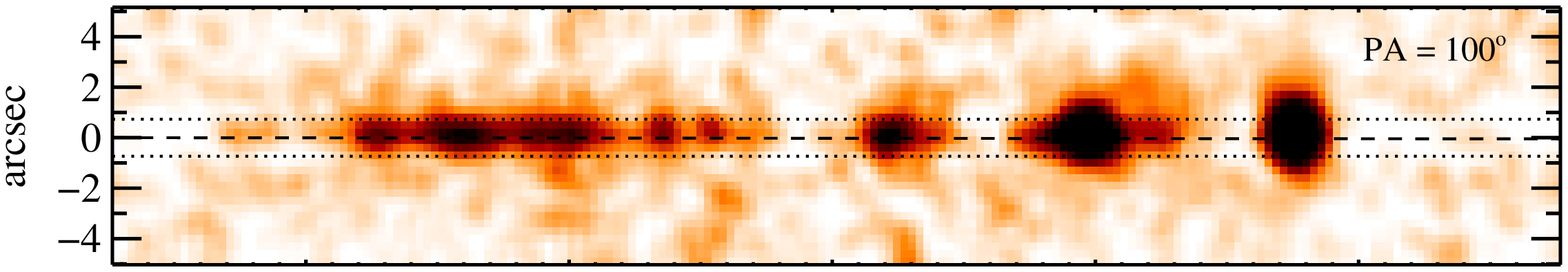}\\
\includegraphics[angle=-270, width=0.95\linewidth, bb=190 90 303 750,clip=true]{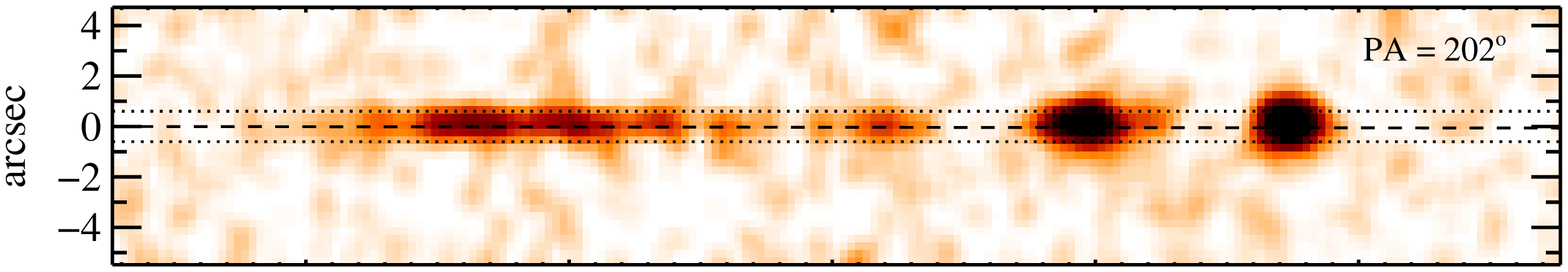}\\
\includegraphics[angle=-270, width=0.95\linewidth, bb=190 90 303 750,clip=true]{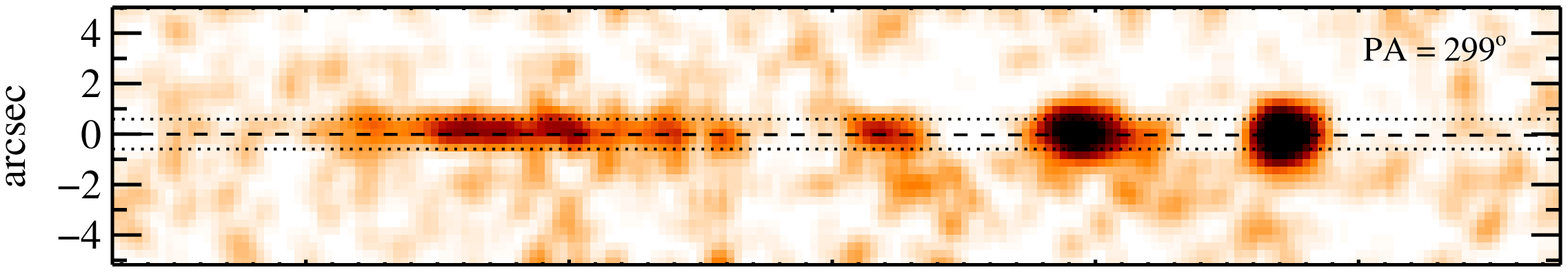}\\
\includegraphics[angle=-270, width=0.95\linewidth, bb=150 90 303 750,clip=true]{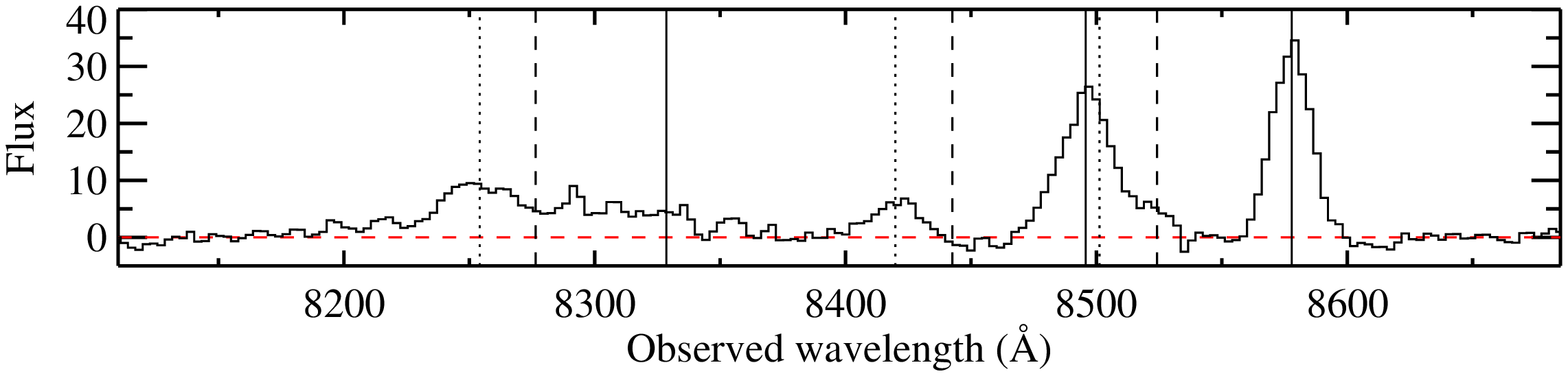}\\
\end{tabular}
\caption{The 2D spectra of J0927$+$2943, in the H$\beta$ and O\,{\sc iii} region, 
after subtraction of the QSO continuum are shown in the top four panels. The dashed 
line shows the centre of the trace used to remove the QSO continuum emission. 
The dotted lines mark the FWHM of the trace. The position angle (PA) of the slit 
is also given in each panel. The bottom panel gives the extracted 1D spectrum to enable 
identifications of different features in the 2D spectra. The vertical lines are as in Fig.~\ref{fig2}.}
\label{fig3}
\end{figure}

Next we perform a 2D spectral analysis. We assume the flux perpendicular to the dispersion 
axis (avoiding the pixels with emission lines) to distribute like a Gaussian around the 
central pixel (i.e. the spectral PSF is assumed to be Gaussian). The continuum flux along the 
dispersion axis is approximated with a lower order polynomial.  As the wavelength range considered is 
very small we use a single FWHM for the spectral PSF. In addition to this, the unblended [O\,{\sc iii}] 
lines (i.e. [O\,{\sc iii}]$\lambda$5008 of the blue system: $L_b$ and [O\,{\sc iii}]$\lambda$4960 of 
the red system: $L_r$) are fitted with 2D Gaussians, using an IDL code based on MPFIT \citep{Markwardt09} 
which performs $\chi^2$-minimisation by applying the Levenberg-Marquart technique.

Two dimensional spectra taken at different position angles of the slit after removing the QSO 
continuum and the background light are shown in the top four panels of Fig.~\ref{fig3}.  The long 
dashed line shows the centre of the trace and dotted lines
mark the FWHM of the spectral point spread function.

The results of the 2D analysis of the data are summarised in Table~\ref{table1}. The first column
in this table gives the position angle. The Gaussian FWHM in the spatial direction 
for the trace and the two unblended emission lines are listed in the next three columns.

From Table~\ref{table1} it is clear that the [O\,{\sc iii}]$\lambda$5008 line of the red component
(i.e L$_r$) has FWHM that is consistently higher ($\ge 4 \sigma$)  than the FWHM of the 
trace.
{Deconvolving from the spectral PSF (as obtained from the FWHM of the QSO trace), this gives 
the physical extent of the emitting region FWHM${_e} =\sqrt{{\rm FWHM(L}_r)^2-{\rm FWHM(QSO)}^2} 
\approx 1.2''\pm0.2''$.
At z$\simeq$0.7, 1.2'' corresponds to a physical extent 
of 8.4~kpc for a flat universe with $\Omega_\Lambda$ = 0.73, $\Omega_m$ = 0.27  and $H_0=73$~\kms\,Mpc$^{-1}$ 
\citep{Komatsu09}.
The FWHM of the [O\,{\sc iii}]$\lambda$4960 line of blue component (L$_b$) is similar 
to that of the trace.  This is consistent with the line emission being unresolved.

The last two entries in Table~\ref{table1} give the relative spatial shift of the emission line
centroids to that of the continuum. It is clear that we see the maximum deviation of the emission 
line centroids for the position angle 299$^{\circ}$. 
In particular L$_r$ is shifted by 0.19$\pm$0.02 arcsec from the quasar trace.
The shift is confirmed by nearly the opposite value for PA=100$^\circ$ while such shifts 
are not seen in other position angles.
The small error is the reflection of the fact that the shift is consistently
seen in all the individual exposures. We also note a $2\sigma$ shift (0.10$\pm$0.05 arcsec) 
for L$_b$ with respect to the trace for PA=299$^{\circ}$. This is probably not statistically significant as 
we do not see any shift in the spectrum taken with PA=100$^\circ$.

Thus we conclude that our observations provide a tentative evidence of the gas associated with
the red component having a projected separation of $\sim 1$~kpc from the quasar. 
This can be easily tested either with direct imaging using HST or repeating 
the same exercise with narrower slit under good seeing conditions.

\section{Discussions}

We report the analysis of long-slit spectroscopic observations of  J0927$+$2943. Comparing our
extracted 1D spectrum with the SDSS spectrum, obtained 4 years before, 
we place a 3$\sigma$ constraint on the acceleration between
the red and blue component to be less than 32 \kms yr$^{-1}$. This is a factor 3 smaller
than the one expected for the binary black hole model \citep{Bogdanovic09}.
However, this alone could not rule out the binary black hole model but rather tightens the 
constraints on the orbital parameters \citep[see][]{Shields09a}. 
Moreover one of the
directly testable predictions of this model is the compact sizes (sub-parsec scale) 
of the emitting regions \citep{Dotti09}. Here we show that the [O\,{\sc iii}] emission from
the red component originates from an extended region of size $\sim$~8\,kpc.  This 
observation probably rules out the binary black hole model for J0927$+$2944.

In the frame work of recoil model with maximally spinning holes we expect the
maximum possible kick of $\sim$4000~\kms \citep{Campanelli07a}.
The extended emission from the
red component can be understood in this model as an effect of photoionisation 
by the accretion disc emission associated with the recoiling black hole. Off centred
emissions are also expected in these models \citep[see][]{Haehnelt06, Loeb07, Guedes09}. 
Thus extended [O\,{\sc iii}] emission from the 
red component or the slight offset we found for this emission with respect to the
QSO trace alone can not rule out the recoiling black hole scenario. 

\citet{Heckman09} proposed that J0927$+$2943 could be a high redshift analog of 
NGC\,1275.
Based on a simple model of in-falling gas photoionised by the QSO continuum, \citeauthor{Heckman09} 
suggested that the observed emission lines could be produced by 
a gas of density 300 cm$^{-3}$ at a distance of 8~kpc from the QSO with a projected 
area of 12 kpc$^2$.  
The extent of the gas we find
($\sim 8$~kpc) for the red component is consistent with \citeauthor{Heckman09}'s simple picture. 
However we wish to point out that  in the case of NGC 1275,  21-cm and X-ray absorption is seen at
the higher redshift suggesting the in-falling gas is in between us and the continuum source
\citep{DeYoung73}. In their calculation, \citeauthor{Heckman09} consider $N$(H) that 
will be optically thick to Lyman continuum radiation. Such gas is also expected to produce
Mg\,{\sc ii} absorption if the in-falling gas is well aligned with the QSO. 
In the SDSS spectrum we do not detect any Mg\,{\sc ii} absorption. 
However, detailed photoionisation modelling is needed to rule out the in falling gas model based on the absence of
Mg\,{\sc ii} absorption.

Our observations confirm the [O\,{\sc iii}]$\lambda$5008  from \zem = 0.7028 reported by
\citet{Shields09a}. To explain the three redshifted emission lines, \citet{Shields09a} proposed a hypothesis
in which different emission components originate from a chance alignment of galaxies that
are part of a massive cluster.

 However there is no clear indication of J0927$+$2943
residing in the centre of a galaxy cluster \citep[see][]{Decarli09}.
In the recoil models this third emission line component 
has to come from the unbound gas that got kicked also with the
black hole. 
Future deep observations under better seeing 
conditions are needed to provide a strong constraint on the spatial extent of the third system. 

\section{Acknowledgements}
{We wish to acknowledge the IUCAA/IGO staff for their support during our observations. 
We thank K. Subramanian and P. Petitjean for useful comments on the manuscript.
We thank Prof. Shyam Tandon for helping us to solve various technical issues related
to the IFOSC data reduction. MV gratefully acknowledges University Grants Commission (UGC),
INDIA, for support through RFSMS Scheme and  IUCAA for hospitality, where most of this work was done.
PN acknowledges support from the french Ministry of European and Foreign Affairs.
}

\def\aj{AJ}%
\def\araa{ARA\&A}%
\def\apj{ApJ}%
\def\apjl{ApJ}%
\def\apjs{ApJS}%
\def\apss{Ap\&SS}%
\def\aap{A\&A}%
\def\aapr{A\&A~Rev.}%
\def\aaps{A\&AS}%
\def\baas{BAAS}%
\def\mnras{MNRAS}%
\def\memras{MmRAS}%
\def\pra{Phys.~Rev.~A}%
\def\prb{Phys.~Rev.~B}%
\def\prc{Phys.~Rev.~C}%
\def\prd{Phys.~Rev.~D}%
\def\pre{Phys.~Rev.~E}%
\def\prl{Phys.~Rev.~Lett.}%
\def\pasp{PASP}%
\def\nat{Nature}%
\def\aplett{Astrophys.~Lett.}%
\def\apspr{Astrophys.~Space~Phys.~Res.}%
\let\astap=\aap
\let\apjlett=\apjl
\let\apjsupp=\apjs
\let\applopt=\ao
\bibliographystyle{mn}

\bibliography{0927}

\begin{thebibliography}{21}
\expandafter\ifx\csname natexlab\endcsname\relax\def\natexlab#1{#1}\fi

\bibitem[{{Bogdanovi{\'c}} {et~al.}(2009){Bogdanovi{\'c}}, {Eracleous}, \&
  {Sigurdsson}}]{Bogdanovic09}
{Bogdanovi{\'c}}, T., {Eracleous}, M., \& {Sigurdsson}, S., 2009, \apj, 697,
  288

\bibitem[{{Boroson} \& {Lauer}(2009)}]{Boroson09}
{Boroson}, T.~A. \& {Lauer}, T.~R., 2009, \nat, 458, 53

\bibitem[{{Campanelli} {et~al.}(2007{\natexlab{a}}){Campanelli}, {Lousto},
  {Zlochower}, \& {Merritt}}]{Campanelli07a}
{Campanelli}, M., {Lousto}, C., {Zlochower}, Y., \& {Merritt}, D.,
  2007{\natexlab{a}}, \apjl, 659, L5

\bibitem[{{Campanelli} {et~al.}(2007{\natexlab{b}}){Campanelli}, {Lousto},
  {Zlochower}, \& {Merritt}}]{Campanelli07b}
{Campanelli}, M., {Lousto}, C.~O., {Zlochower}, Y., \& {Merritt}, D.,
  2007{\natexlab{b}}, \prl, 98, 231102

\bibitem[{{Conselice} {et~al.}(2001){Conselice}, {Gallagher}, \&
  {Wyse}}]{Conselice01}
{Conselice}, C.~J., {Gallagher}, III, J.~S., \& {Wyse}, R.~F.~G., 2001, \aj,
  122, 2281

\bibitem[{{Dain} {et~al.}(2008){Dain}, {Lousto}, \& {Zlochower}}]{Dain08}
{Dain}, S., {Lousto}, C.~O., \& {Zlochower}, Y., 2008, \prd, 78, 024039

\bibitem[{{De Young} {et~al.}(1973){De Young}, {Roberts}, \&
  {Saslaw}}]{DeYoung73}
{De Young}, D.~S., {Roberts}, M.~S., \& {Saslaw}, W.~C., 1973, \apj, 185, 809

\bibitem[{{Decarli} {et~al.}(2009){Decarli}, {Reynolds}, \&
  {Dotti}}]{Decarli09}
{Decarli}, R., {Reynolds}, M.~T., \& {Dotti}, M., 2009, \mnras, 397, 458

\bibitem[{{Dotti} \& {Volonteri}(2009)}]{Dotti09}
{Dotti}, M. \& {Volonteri}, M., 2009, in Bulletin of the American Astronomical
  Society, Vol.~41, Bulletin of the American Astronomical Society, p. 385

\bibitem[{{Gonz{\'a}lez} {et~al.}(2007){Gonz{\'a}lez}, {Hannam}, {Sperhake},
  {Br{\"u}gmann}, \& {Husa}}]{Gonzalez07}
{Gonz{\'a}lez}, J.~A., {Hannam}, M., {Sperhake}, U., {Br{\"u}gmann}, B., \&
  {Husa}, S., 2007, \prl, 98, 231101

\bibitem[{{Guedes} {et~al.}(2009){Guedes}, {Madau}, {Kuhlen}, {Diemand}, \&
  {Zemp}}]{Guedes09}
{Guedes}, J., {Madau}, P., {Kuhlen}, M., {Diemand}, J., \& {Zemp}, M., 2009,
  \apj, accepted [ArXiv:0907.0892]

\bibitem[{{Haehnelt} {et~al.}(2006){Haehnelt}, {Davies}, \&
  {Rees}}]{Haehnelt06}
{Haehnelt}, M.~G., {Davies}, M.~B., \& {Rees}, M.~J., 2006, \mnras, 366, L22

\bibitem[{{Heckman} {et~al.}(2009){Heckman}, {Krolik}, {Moran}, {Schnittman},
  \& {Gezari}}]{Heckman09}
{Heckman}, T.~M., {Krolik}, J.~H., {Moran}, S.~M., {Schnittman}, J., \&
  {Gezari}, S., 2009, \apj, 695, 363

\bibitem[{{Komatsu} {et~al.}(2009){Komatsu}, {Dunkley}, {Nolta}, {Bennett},
  {Gold}, {Hinshaw}, {Jarosik}, {Larson}, {Limon}, {Page}, {Spergel},
  {Halpern}, {Hill}, {Kogut}, {Meyer}, {Tucker}, {Weiland}, {Wollack}, \&
  {Wright}}]{Komatsu09}
{Komatsu}, E., {Dunkley}, J., {Nolta}, M.~R., {et~al.}, 2009, \apjs, 180, 330

\bibitem[{{Komossa} {et~al.}(2008){Komossa}, {Zhou}, \& {Lu}}]{Komossa08}
{Komossa}, S., {Zhou}, H., \& {Lu}, H., 2008, \apjl, 678, L81

\bibitem[{{Loeb}(2007)}]{Loeb07}
{Loeb}, A., 2007, \prl, 99, 041103

\bibitem[{{Markwardt}(2009)}]{Markwardt09}
{Markwardt}, C.~B., 2009, ArXiv e-prints 0902.2850

\bibitem[{{Shields} {et~al.}(2009{\natexlab{a}}){Shields}, {Bonning}, \&
  {Salviander}}]{Shields09a}
{Shields}, G.~A., {Bonning}, E.~W., \& {Salviander}, S., 2009{\natexlab{a}},
  \apj, 696, 1367

\bibitem[{{Shields} {et~al.}(2009{\natexlab{b}}){Shields}, {Rosario}, {Smith},
  {Bonning}, {Salviander}, {Kalirai}, {Strickler}, {Ramirez-Ruiz}, {Dutton},
  {Treu}, \& {Marshall}}]{Shields09b}
{Shields}, G.~A., {Rosario}, D.~J., {Smith}, K.~L., {et~al.},
  2009{\natexlab{b}}, \apj, submitted [ArXiv:0907.3470]

\bibitem[{{Tichy} \& {Marronetti}(2007)}]{Tichy07}
{Tichy}, W. \& {Marronetti}, P., 2007, \prd, 76, 061502

\bibitem[{{Vanden Berk} {et~al.}(2001){Vanden Berk}, {Richards}, {Bauer},
  {Strauss}, {Schneider}, {Heckman}, {York}, {Hall}, {Fan}, {Knapp},
  {Anderson}, {Annis}, {Bahcall}, {Bernardi}, {Briggs}, {Brinkmann}, {Brunner},
  {Burles}, {Carey}, {Castander}, {Connolly}, {Crocker}, {Csabai}, {Doi},
  {Finkbeiner}, {Friedman}, {Frieman}, {Fukugita}, {Gunn}, {Hennessy},
  {Ivezi{\'c}}, {Kent}, {Kunszt}, {Lamb}, {Leger}, {Long}, {Loveday}, {Lupton},
  {Meiksin}, {Merelli}, {Munn}, {Newberg}, {Newcomb}, {Nichol}, {Owen}, {Pier},
  {Pope}, {Rockosi}, {Schlegel}, {Siegmund}, {Smee}, {Snir}, {Stoughton},
  {Stubbs}, {SubbaRao}, {Szalay}, {Szokoly}, {Tremonti}, {Uomoto}, {Waddell},
  {Yanny}, \& {Zheng}}]{VandenBerk01}
{Vanden Berk}, D.~E., {Richards}, G.~T., {Bauer}, A., {et~al.}, 2001, \aj, 122,
  549

\end{thebibliography}
\label{lastpage}

\end{document}